\documentclass[
 reprint,
 amsmath,amssymb,
 aps,superscriptaddress,
]{revtex4-2}

\usepackage{graphicx}
\usepackage{dcolumn}
\usepackage{bm}
\usepackage{physics}
\usepackage{braket}
\setcitestyle{super}

\begin{document}

\title{
Cavity-mediated coherence protection and one-axis twisting for spins in solids}

\author{Rikuto~Fukumori}
\thanks{These authors contributed equally to this work.}
\affiliation{Thomas J. Watson, Sr., Laboratory of Applied Physics, California Institute of Technology, Pasadena, CA 91125, USA}
\affiliation{Kavli Nanoscience Institute, California Institute of Technology, Pasadena, CA 91125, USA}
\affiliation{Institute for Quantum Information and Matter, California Institute of Technology, Pasadena, CA 91125, USA}

\author{Chengyi~Luo}
\thanks{These authors contributed equally to this work.}
\affiliation{Thomas J. Watson, Sr., Laboratory of Applied Physics, California Institute of Technology, Pasadena, CA 91125, USA}
\affiliation{Kavli Nanoscience Institute, California Institute of Technology, Pasadena, CA 91125, USA}
\affiliation{Institute for Quantum Information and Matter, California Institute of Technology, Pasadena, CA 91125, USA}

\author{Alexey~Tiranov}
\affiliation{Chimie ParisTech, PSL University, CNRS, Institut de Recherche de Chimie Paris, 75005 Paris, France}

\author{Karolina~Waszkowska}
\affiliation{Chimie ParisTech, PSL University, CNRS, Institut de Recherche de Chimie Paris, 75005 Paris, France}

\author{Philippe~Goldner}
\affiliation{Chimie ParisTech, PSL University, CNRS, Institut de Recherche de Chimie Paris, 75005 Paris, France}

\author{Andrei~Faraon}
\email{faraon@caltech.edu}
\affiliation{Thomas J. Watson, Sr., Laboratory of Applied Physics, California Institute of Technology, Pasadena, CA 91125, USA}
\affiliation{Kavli Nanoscience Institute, California Institute of Technology, Pasadena, CA 91125, USA}
\affiliation{Institute for Quantum Information and Matter, California Institute of Technology, Pasadena, CA 91125, USA}

\date{March 16, 2026}

\begin{abstract}
    Long-range interactions between emitters give rise to collective phenomena, including superradiance, spin squeezing, and coherence protection, that are important to both fundamental physics and quantum technologies. Despite progress in cold atoms, coherent cavity-mediated all-to-all interactions have not yet been realized in a solid-state ensemble. Here we demonstrate such interactions in a $^{171}$Yb$^{3+}$:CaWO$_4$ crystal coupled to a microwave resonator, observing superradiant emission on resonance and unitary one-axis twisting dynamics in the dispersive regime. The same interaction also opens a many-body energy gap that suppresses inhomogeneous dephasing, extending the ensemble Ramsey coherence time from tens of microseconds to milliseconds without decoupling pulses. These results establish a solid-state platform for collective many-body physics with direct implications for quantum technologies. Specifically, the observed one-axis twisting dynamics opens a path towards spin squeezing for entanglement-enhanced quantum metrology, and the extended coherence due to gap-protection is relevant for both microwave photon storage and precision measurement.
\end{abstract}

\maketitle

The collective behavior of interacting many-body quantum systems underpins some of the most important developments in quantum science. When long-range couplings connect the dynamics of many particles through shared modes, new phenomena emerge: superradiant emission\cite{Dicke1954,GrossHaroche1982}, one-axis twisting (OAT) dynamics \cite{KitagawaUeda1993,Wineland1992}, and the emergence of a many-body energy gap for coherence protection\cite{Rey2008,Norcia2018,Davis2020,Thompson2024Science,NiuPRL2025}. Cold atoms and trapped ions have been the primary experimental platforms for these collective effects\cite{Esslinger2010Nature,Bohnet2012,Thompson2016SA,Leroux2010,SchleierSmith2010,ThompsonPRL2016,Kasevich2016Nature,Norcia2018,Davis2020,Thompson2024Science,NiuPRL2025,Bohnet2016,Britton2012,MonroeRMP2021}. Realizing the same cavity-mediated interactions in a solid-state spin ensemble would access a qualitatively different regime: crystalline hosts contain vast spin ensembles ($\sim\!10^{15}$ spins in a millimeter-scale crystal) far larger than those in typical atomic implementations, operate natively in the millikelvin, microwave-frequency regime of superconducting circuits, and offer simultaneous optical access for microwave-optical transduction. Fundamentally, solid-state spin ensembles are subject to intrinsic crystalline disorder that limits the Ramsey coherence time T$_2^*$. Although Hahn echo can refocus this static disorder and extend coherence to T$_2$, many applications remain fundamentally limited by T$_2^*$. For sensing of static or slowly-varying signals, T$_2^*$ bounds the interrogation time as the echo sequence refocuses both the disorder and the signal. For quantum memories, a longer T$_2^*$ would reduce the need for dynamical decoupling and the associated pulse errors and added noise\cite{GREZES2016693}.

Strong collective coupling of solid-state spin ensembles to microwave resonators is well established\cite{Putz2014,KuboPRL2010,AmsussPRL2011,SchusterPRL2010,RanjanPRL2013,CreedonPRB2015,ProbstPRL2013,Faraon2025NatPhys,Wang2022PRA,Dold2019PRA,RosePRX2017,Zollitsch2015APL,Weichselbaumer2020PRL}, and collective radiative effects including superradiant emission have been observed\cite{KerstenPRL2023,Angerer2018,Lei2023Nature}. Yet demonstrations of coherent many-body dynamics in these systems have been limited to finite-range dipolar interactions\cite{ChoiPRL2017,DavisNatPhys2023,Yao2025arXiv,Lei2025} or cavity-mediated effects in systems with a limited number of qubits or emitters such as superconducting qubits\cite{Zhu2019Science} or diamond color centers\cite{Evans2018}. Unitary cavity-mediated all-to-all interactions have not yet been realized in a solid-state spin ensemble.

Here we report the observation of cavity-mediated spin-exchange interactions in a solid-state system and demonstrate three key manifestations of the resulting collective dynamics: superradiance, OAT dynamics, and gap-protected coherence. Using a $^{171}$Yb$^{3+}$:CaWO$_4$ crystal coupled to a 3D microwave resonator, we access two complementary regimes by tuning the spin--resonator detuning. On resonance, the ensemble exhibits superradiant emission with the characteristic $N^2$ scaling. In the dispersive regime, the shared cavity mode mediates unitary all-to-all spin-exchange interactions that generate OAT dynamics. The same interaction also opens a many-body energy gap between collective states of different symmetry, suppressing inhomogeneous dephasing and extending the Ramsey T$_2^*$ from $52~\mu\mathrm{s}$ to $3.3~\mathrm{ms}$.

\begin{figure*}
    \centering
    \includegraphics[width=\linewidth]{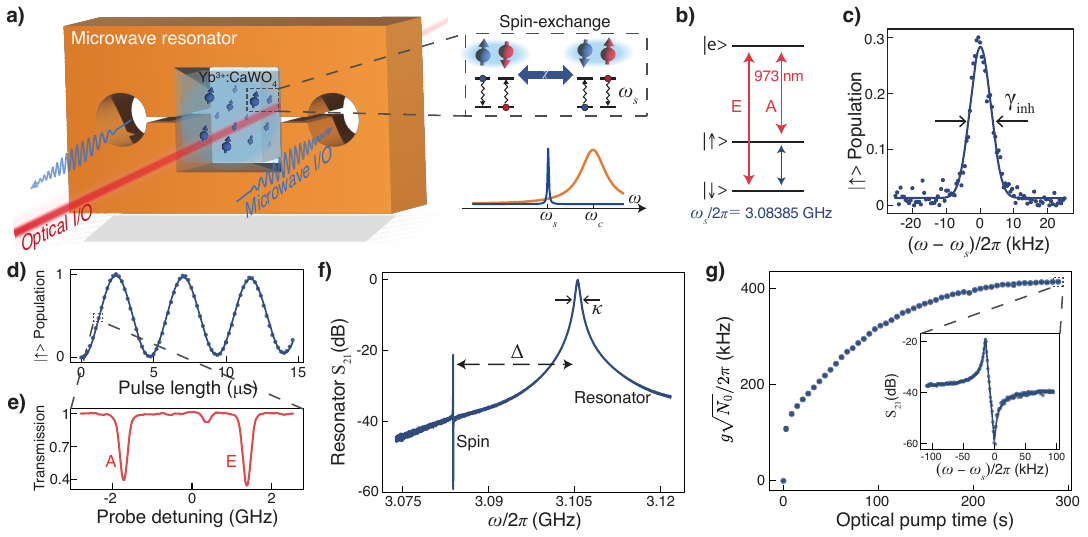}
    \caption{\textbf{System overview.} \textbf{(a)} Cutaway of a loop-gap microwave resonator hosting a millimeter-scale $^{171}$Yb$^{3+}$:CaWO$_4$ crystal. Red beam: optical input/output (I/O) at $\sim973~\mathrm{nm}$ for initialization and readout; spin excitation and resonator transmission measurements are done via antennas at the microwave I/O ports on the outer loops. Right: schematic of cavity-mediated spin-exchange interactions. \textbf{(b)} Simplified energy level diagram shows the spin manifold {$\lvert\downarrow\rangle$,$\lvert\uparrow\rangle$} split by $\omega_s/2\pi=3.08385$~GHz and two optical transitions (A,E) to an excited state $\lvert\text{e}\rangle$ used for pumping and absorption-based population readout. \textbf{(c)} Spin spectroscopy obtained by sweeping the frequency of a weak $2$~ms excitation pulse; a Gaussian fit gives  $\gamma_\mathrm{inh}/2\pi<5$~kHz. \textbf{(d)} Ensemble Rabi oscillations. Normalized $\lvert\uparrow\rangle$ population versus microwave excitation pulse length, driven through the resonator input and read out via optical absorption. \textbf{(e)} Optical absorption spectrum after a $\pi/2$ pulse. A,E transition depths give the $\lvert\uparrow\rangle$,$\lvert\downarrow\rangle$ populations, respectively. \textbf{(f)} Weak-probe transmission $S_{21}(\omega)$ of the resonator used in the dispersive measurements (with linewidth $\kappa/2\pi=660$~kHz) and the spin ensemble response. Here the spin--resonator detuning is $\Delta/2\pi=22$~MHz. \textbf{(g)} Extracted ensemble coupling strength ($g\sqrt{N_0}$) versus optical pumping time. Inset: representative $S_{21}(\omega)$ and fit to model near the spin resonance (SI).}
    \label{fig:fig1}
\end{figure*}

\section*{System overview}
Our system comprises a millimeter-scale $^{171}$Yb$^{3+}$:CaWO$_4$ crystal containing $\sim10^{15}$ $^{171}$Yb$^{3+}$ ions (Fig.~\ref{fig:fig1}a). The ions support optical transitions near $973$~nm and a hyperfine ground-state manifold with a microwave transition at $\omega_s/2\pi = 3.08385$~GHz between states $\lvert\uparrow\rangle$ and $\lvert\downarrow\rangle$ (Fig.~\ref{fig:fig1}b). At zero magnetic field, this microwave transition lies at a ZEFOZ (zero first-order Zeeman) point, yielding spin-echo T$_2>150$~ms\cite{Tiranov2025} and an inhomogeneous linewidth $\gamma_{\text{inh}}/2\pi<5$~kHz (Fig.~\ref{fig:fig1}c). This combination of long intrinsic coherence, a substantial magnetic dipole moment inherited from the electronic component of this hybridized electron–nuclear transition (SI), and high ion density enables strong coupling to microwave fields while maintaining first-order insensitivity to magnetic noise, a regime difficult to access in other solid-state spin systems.

To couple the ensemble to a common microwave mode, the crystal is mounted inside a loop-gap resonator that confines the oscillating magnetic field. The coupled system is described by the Tavis--Cummings Hamiltonian\cite{TavisCummings1968} in the rotating frame of $\omega_s$:
\begin{equation}
    H=\Delta \hat{a}^{\dagger}\hat{a} + \frac{1}{2}\sum_{j=1}^{N}\delta_j\hat{\sigma}_{z,j}+g\sum_{j=1}^N(\hat{a}^{\dagger}\hat{\sigma}_{-,j}+\hat{\sigma}_{+,j}\hat{a}),
    \label{eq:Hfull}
\end{equation}
where $\hat{a}$ is the resonator mode operator, $\hat{\sigma}_{(\pm,z),j}$ are the Pauli operators for the $j$th spin with frequency offset $\delta_j$ from $\omega_s$, $\Delta=\omega_c-\omega_s$ is the resonator--spin detuning, and $g$ is the spin--cavity coupling strength for a single ion, which we approximate as uniform across the crystal given the highly homogeneous resonator field over the sample volume (SI). This spatial uniformity enables coherent global control of the spin ensemble. To realize this control, we optically pump the spins into $\lvert\downarrow\rangle$, apply microwave pulses of variable duration through the resonator to drive the spins, and read out the resulting $\lvert\uparrow\rangle$ and $\lvert\downarrow\rangle$ populations via optical absorption spectroscopy on the A and E transitions (Fig.~\ref{fig:fig1}e). The clean ensemble Rabi oscillations in Fig.~\ref{fig:fig1}d establish that this control acts coherently across the ensemble and underpins the experiments that follow.

\begin{figure*}
    \centering
    \includegraphics[width=0.92\linewidth]{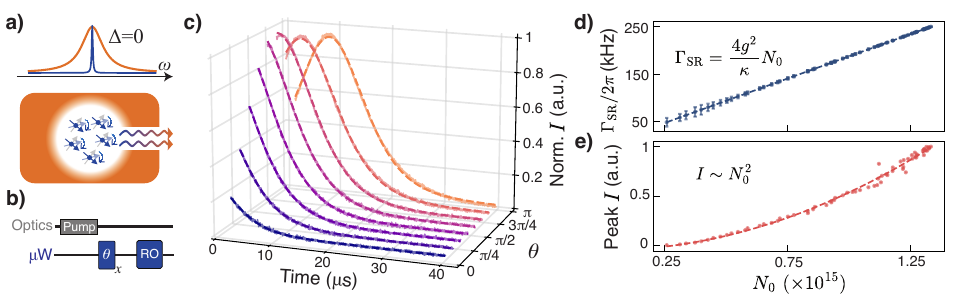}
    \caption{\textbf{Superradiant emission.} \textbf{(a)} The spin and resonator are approximately on resonance ($\Delta\approx0$), resulting in collectively enhanced emission through the resonator output. \textbf{(b)} Pulse sequence. Optical initialization $\to$ microwave rotation to polar angle $\theta$ $\to$ readout of the microwave emission from the resonator. \textbf{(c)} Emission dynamics (normalized intensity) versus initial polar angle $\theta$, with simulations (dashed) overlaid. Near inversion ($\theta\approx\pi$), the characteristic delayed superradiant burst is visible. \textbf{(d)} Superradiant emission rate $\Gamma_{\mathrm{SR}}$ versus $N_0$ at $\theta = \pi/6$. $\Gamma_{\mathrm{SR}}$ is extracted by fitting the emission traces to numerical simulations (SI). The rate scales linearly with $N_0$ (dashed line), consistent with the expected collective enhancement $\Gamma_{\mathrm{SR}} \propto N_0$. \textbf{(e)} Peak emission intensity versus $N_0$ at near-inversion ($\theta\approx\pi$), fit to a quadratic dependence $I \propto N_0^2$ (dashed line), confirming the characteristic $N^2$ scaling of Dicke superradiance.}
    \label{fig:fig2}
\end{figure*}

We next characterize the coupled spin--resonator system via the resonator transmission $S_{21}(\omega)$ (Fig.~\ref{fig:fig1}f). From the cavity response over the full transmission spectrum, we determine the spin--resonator detuning $\Delta$ and linewidth $\kappa$. With $\Delta$ and $\kappa$ fixed, we then fit the narrow feature near $\omega_s$ to a model derived from Eq.~(\ref{eq:Hfull}) (SI), to extract the collective coupling rate $g\sqrt{N_0}$. Combining this fitted collective coupling with the single-ion value $g/2\pi=15$~mHz obtained from simulations of the resonator field (SI), we extract $N_0$, the effective number of polarized spins. Throughout this work, $N_0$ denotes this calibrated effective spin number inferred from the fitted coupling spectrum. We note that while the optical beam covers only $\sim1\%$ of the crystal face, spin diffusion redistributes polarization throughout the bulk, leading to a larger $N_0$ with longer optical pumping duration (Fig.~\ref{fig:fig1}g).

Because intrinsic spin relaxation (T$_1\gg1$~s) and homogeneous dephasing (T$_2>150$~ms) are negligible on the timescales studied here, the effective dynamics are set primarily by the resonator detuning relative to its linewidth. When the resonator is tuned near resonance with the spin transition ($\Delta\approx 0$), collective dissipation dominates and the ensemble superradiates. In the dispersive regime ($|\Delta|\gg\kappa$), this dissipation is quadratically suppressed and unitary spin-exchange interactions emerge. We explore both regimes in the following sections, using two loop-gap resonators with linewidths optimized for each case.

\section*{Superradiant emission}
To access collective dissipation, we first use a loop-gap resonator with a linewidth $\kappa/2\pi = 4.8~\mathrm{MHz}$, tuned approximately on resonance with the spin transition ($\Delta\approx 0$; Fig.~\ref{fig:fig2}a). In this regime, the cavity linewidth exceeds all other relevant frequency scales ($\kappa\gg g\sqrt{N_0},\gamma_\text{inh},|\Delta|$), and the ensemble radiates collectively through the shared resonator mode at a superradiant emission rate
\begin{equation}
\Gamma_{\mathrm{SR}}\approx\frac{4 g^2 N_0}{\kappa},
\label{gamma_sr}
\end{equation}
collectively enhanced by a factor of $N_0$ over the resonant single-spin Purcell rate. For our parameters, $\Gamma_\text{SR}$ reaches hundreds of kilohertz, vastly exceeding the spin-lattice relaxation rate (T$_1$ of minutes to hours \cite{Tiranov2025}), such that depolarization during the experiment is dominated by cavity-mediated collective emission. To measure the superradiant emission, we first optically pump the spin ensemble into $\lvert\downarrow\rangle$, seed the superradiance by applying a fast global microwave rotation to prepare a coherent spin state at polar angle $\theta$ from the south pole, and then record the microwave field emitted from the resonator output port (Fig.~\ref{fig:fig2}b).

\begin{figure*}
    \centering
    \includegraphics[width=\linewidth]{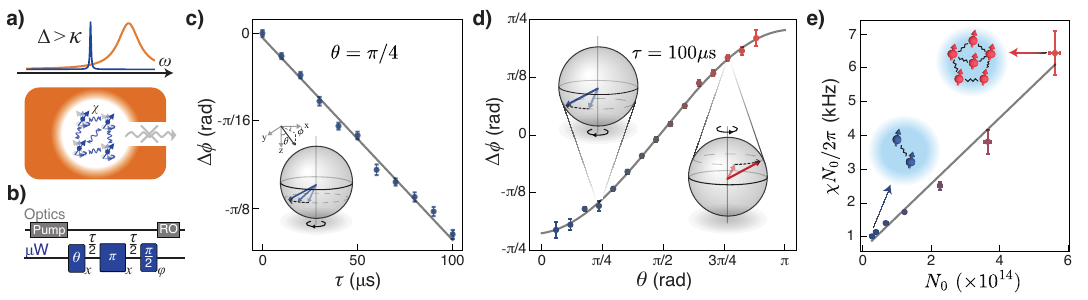}
    \caption{\textbf{One-axis twisting dynamics.} \textbf{(a)} The resonator is detuned from the spin ($|\Delta|\gg\kappa$), resulting in unitary cavity-mediated interactions while suppressing superradiant emission. \textbf{(b)} Pulse sequence. Optical initialization is followed by a spin Hahn-echo sequence with initial rotation $\theta$. The phase of the final $\pi/2$ pulse ($\varphi$) is varied to extract the phase shift induced by OAT dynamics. \textbf{(c)} Evolution of the Bloch-vector azimuthal angle $\Delta\phi$ versus $\tau$ at $\theta=\pi/4$. The slope of the overlaid linear fit gives $\chi N_0 \cos(\pi/4)$. \textbf{(d)} Evolution of $\Delta\phi$ vs.\ $\theta$ at $\tau=100~\mu$s. Data is fit to $\Delta\phi=\chi N_0\tau\cos{\theta}$. \textbf{(e)} Collective interaction rate $\chi N_0$ vs. $N_0$ at $\theta=\pi/4$. For each optical pumping time, $\chi N_0$ (y-axis) is extracted as in (c) from the OAT phase shift as a function of time, and the $N_0$ (x-axis) is extracted from resonator $S_{21}(\omega)$ with the same method shown in Fig.~\ref{fig:fig1}g. The linear dependence $\chi N_0 \propto N_0$ (dashed line) confirms the expected scaling, with the slope yielding the single-spin interaction strength $\chi$. The free parameter of this fit is $\Delta/2\pi=23(2)$ MHz, matching the independently measured $\Delta/2\pi=22$ MHz from $S_{21}(\omega)$.}
    \label{fig:fig3}
\end{figure*}

The time-resolved emission traces (Fig.~2c) display the hallmark features of Dicke superradiance over the full range of initial polar angles. For initial states below the equator ($\theta<\pi/2$), the emission intensity is largest at the start and decays monotonically as the collective Bloch vector tips toward the south pole. For initial states above the equator ($\theta>\pi/2$), the dynamics change qualitatively, exhibiting the characteristic delayed superradiant burst, with the peak emission getting delayed further as the initial state approaches full inversion. This behavior can be understood from the motion of the collective Bloch vector: the emission rate first increase as the collective Bloch vector tips toward the equator, reaches a peak at the equator, and then decays as the ensemble relaxes to the ground state $\lvert\downarrow\rangle$. Across the full range of initial states, the measured traces are in quantitative agreement with numerical simulations (dashed lines in Fig.~2c) using a single parameter set.

To quantitatively verify the collective origin of the emission, we tune $N_0$ by varying the optical pumping duration and extract $\Gamma_\text{SR}$ from the emission dynamics at a fixed polar angle $\theta=\pi/6$ as the initial state (Fig.~2d). The measured rate scales linearly with $N_0$, as predicted by Eq.~\ref{gamma_sr}, with a fitted slope that quantitatively agrees with the experimental parameters. We obtain an independent scaling test from the peak emission intensity for an initial state near full inversion ($\theta\approx\pi$; Fig.~2e). Starting near inversion ensures that the emission peak is delayed from the excitation pulse, making it robust against timing errors. The peak intensity scales as $N_0^2$, confirming the characteristic quadratic scaling of Dicke superradiance.

The delayed burst near inversion, the $\theta$-dependent emission dynamics, and the nonlinear $N_0$ scaling are clear signatures of Dicke superradiance in the collective spin--resonator system. These observations evidence the collective nature of the system that underlies the coherent interaction physics explored below.

\section*{One-axis twisting dynamics}
We now detune the resonator from the spin transition to enter the dispersive regime (Fig.~\ref{fig:fig3}a), where the same spin-cavity coupling that produced superradiant emission instead mediates spin-spin interactions. In this limit, adiabatic elimination of the cavity yields the spin-spin interaction described by the effective spin-exchange Hamiltonian
\begin{equation}
    \hat{H}_{\mathrm{eff}} = \chi \hat{J}_+ \hat{J}_-=\chi(\hat{J}^2 - \hat{J}_z^2 + \hat{J}_z),
    \label{eq:Heff}
\end{equation}
where $\hat{J}_{\pm}=\sum_j \hat{\sigma}_{\pm,j}$ and $\hat{J}_{z}=\tfrac12\sum_j \hat{\sigma}_{z,j}$ are the collective spin operators, and $\chi=g^2/\Delta$ is the characteristic cavity-mediated interaction strength (general form in SI). The collective interaction strength $\chi N_0$ scales as $1/|\Delta|$, while the superradiant emission rate $\Gamma_\text{SR}$ scales as $1/\Delta^2$, so at large detuning unitary dynamics dominates over collective dissipation. Within a fixed-$J$ manifold, the $\chi\hat{J}^2$ is constant and the single-particle term $\chi \hat{J}_z$ is negligible in the large-$N$ limit, so the nontrivial dynamics is generated by $\hat{H}_{\text{OAT}}=-\chi\hat{J}_z^2$, which realizes OAT dynamics\cite{KitagawaUeda1993}. At the mean-field level, this induces a precession of the collective Bloch vector about the $z$-axis at a rate proportional to the population imbalance, $-2\chi\langle J_z\rangle$.

The signature of OAT is therefore an accumulated phase shift set by the polar angle of the collective spin state. To probe this, we employ a modified Hahn-echo sequence (Fig.~\ref{fig:fig3}b): $N_0$ spins initialized in $\lvert\downarrow\rangle$ are rotated by angle $\theta$, undergo free evolution for time $\tau$ with a central $\pi$ pulse that cancels static frequency errors, and are read out with a final $\pi/2$ pulse whose phase $\varphi$ is varied to extract the accumulated phase shift $\Delta\phi$\cite{Thompson2025NatPhys}. The echo refocuses inhomogeneous dephasing, isolating the nonlinear $\Delta\phi\propto\cos\theta$ contribution from OAT dynamics. For this experiment, we use a loop-gap resonator with a narrower linewidth $\kappa/2\pi=660$~kHz and a large detuning $\Delta/2\pi=22$~MHz $\gg\kappa$ from the spin transition to suppress superradiance.

\begin{figure*}
    \centering
    \includegraphics[width=0.71\linewidth]{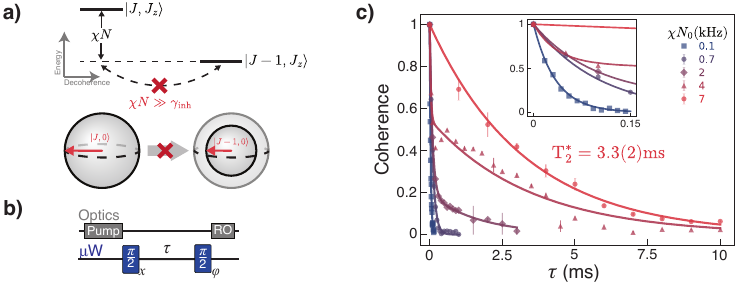}
    \caption{\textbf{Many-body gap protection.}
    \textbf{(a)} The $\chi \hat{J}^2$ term opens a many-body energy gap of size $\chi N_0$ between a collective state $\lvert J,J_z\rangle$ and a less coherent state $\lvert J-1,J_z\rangle$, suppressing inhomogeneous dephasing. In the Bloch sphere representation, this corresponds to a suppression of the shrinkage of the collective Bloch vector.
    \textbf{(b)} Experimental pulse sequence for measuring the Ramsey coherence time. The readout is performed optically in the population basis.
    \textbf{(c)} Ramsey coherence versus delay time $\tau$ for five values of the collective interaction rate $\chi N_0/2\pi \approx 0.1, 0.7, 2, 4, 7$~kHz (blue to red). At low $\chi N_0$, the coherence decays with the disorder-limited Ramsey time $\mathrm{T}_2^*\!=\!52(3)~\mu$s. As $\chi N_0$ exceeds $\gamma_{\mathrm{inh}}$, protection is achieved, reaching $\mathrm{T}_2^*\!=\!3.3(2)$~ms at the highest interaction strength. Solid lines are exponential (or double-exponential) fits. Inset: zoom in to the first $150$~$\mu$s.}
    \label{fig:fig4}
\end{figure*}

We first verify that the phase accumulates linearly in time. Setting $g\sqrt{N_0}/2\pi\sim150$~kHz and $\theta=\pi/4$, we sweep $\tau$ and observe a steady decrease of $\Delta\phi$ (Fig.~\ref{fig:fig3}c). From the measured slope, together with the known $\cos\theta$ dependence, we extract $\chi N_0/2\pi = 1.04(4)~\mathrm{kHz}$. To confirm the population-dependent character that distinguishes OAT from a linear frequency shift, we fix $\tau=100~\mu$s and scan $\theta$ (Fig.~\ref{fig:fig3}d). The phase shift reverses sign for initial states above versus below the equator and vanishes at the equator. A fit to the analytical form $\Delta\phi(\theta)=\chi N_0 \tau \cos(\theta)$ yields $\chi N_0 /2\pi = 1.10(7)$~kHz, consistent with both the time-domain measurement and the expected $g^2N_0/\Delta$. Finally, we vary $N_0$ and correlate $\chi N_0$ extracted from OAT phase accumulation with $N_0$ calibrated independently from resonator $S_{21}(\omega)$ (Fig.~\ref{fig:fig3}e). The self-consistent linear dependence $\chi N_0\propto N_0$ confirms the expected scaling, with a fitted slope consistent with experimental parameters.

The quantitative agreement across time dependence, angular dependence, and $N_0$-scaling constitutes a direct observation of cavity-mediated OAT dynamics in a solid-state ensemble.

\section*{Many-body gap protection}
We now turn to the $\chi \hat{J}^2$ term in the spin-exchange Hamiltonian which opens a many-body energy gap\cite{Rey2008} between states of different total spin quantum number $J$. The gap between adjacent $J$-manifolds is
\begin{equation}
    E_{J}-E_{J-1} =\chi\left[J(J{+}1)-(J{-}1)J\right] = 2\chi J = \chi N_0.
\end{equation}
This gap energetically isolates the fully symmetric manifold ($J=N_0/2$) from states of lower permutational symmetry (Fig.~\ref{fig:fig4}a). At the mean-field level, the nonlinear term $\chi \hat{J}^2\approx 2\chi \langle \hat{J} \rangle \hat{J}$ induces a rotation about the collective Bloch vector, with angular frequency $2\chi \langle \hat{J}\rangle$. This mechanism is analogous to spin locking, but with the locking field generated internally rather than applied externally. When the many-body energy gap exceeds the inhomogeneous linewidth ($\chi N_0 \gtrsim \gamma_{\text{inh}}$), disorder-induced dephasing is suppressed and the ensemble Ramsey coherence time T$_2^*$ is extended beyond the inhomogeneity limit (SI).

We measure the Ramsey coherence decay for five values of $\chi N_0$, spanning the crossover from the dephasing-dominated to the interaction-dominated regime (Fig.~\ref{fig:fig4}b). At low interaction strength ($\chi N_0/2\pi \approx 0.1$~kHz $\ll \gamma_{\mathrm{inh}}/2\pi$), the coherence decays with T$_2^* = 52(3)~\mu\mathrm{s}$, consistent with the independently measured inhomogeneous broadening. As $\chi N_0$ approaches $\gamma_{\mathrm{inh}}$, disorder-induced dephasing is progressively suppressed by many-body gap protection. In the interaction-dominated regime, at the highest interaction strength ($\chi N_0/2\pi \approx 7~\mathrm{kHz} > \gamma_{\mathrm{inh}}/2\pi$), the Ramsey decay is well described by a single exponential with T$_2^* = 3.3(2)~\mathrm{ms}$, corresponding to a $\sim\!65\times$ enhancement over the disorder-limited value. This is consistent with protection of nearly the entire participating ensemble, and the remaining decay is limited by residual superradiant emission (SI), indicating that even longer coherence can be realized with a higher-$Q$ resonator.

\section*{Discussion}
The one-axis twisting dynamics and many-body gap protection demonstrated here constitute, to our knowledge, the first observation of unitary cavity-mediated all-to-all interactions in a solid-state ensemble. Achieving this interaction-dominated regime requires $\Gamma_{\mathrm{SR}} \ll \chi N$ and $\chi N \gtrsim \gamma_{\mathrm{inh}}$. The key ingredients enabling access to this regime are a narrow inhomogeneous linewidth and a large magnetic dipole moment, which together enable strong ion--cavity coupling without sacrificing coherence. This combination is realized in $^{171}$Yb$^{3+}$:CaWO$_4$ at its ZEFOZ point and is difficult to achieve in most other solid-state systems, where broader linewidths typically require larger detunings that reduce the interaction rate.

Several immediate extensions follow from these results. First, the demonstrated OAT dynamics provides a direct path to spin squeezing in the solid state. In the current setup, the achievable squeezing is estimated to be about 2~dB below the standard quantum limit, limited by the collective cooperativity\cite{Rey2018PRL}. This can be substantially improved with superconducting resonators\cite{Faraon2025NatPhys} that combine narrower cavity linewidth with strong collective coupling. Direct observation of metrological gain is currently limited by imperfect state initialization and technical detection resolution. Looking ahead, going beyond mean--field will require better state preparation, improved readout sensitivity\cite{SushkovNatPhys2026}, or interaction-based readout protocols\cite{Davis2016PRL}.

Second, the echo-less extension of T$_2^*$ has direct implications for Ramsey-based sensing and frequency metrology. In such protocols, the interrogation time, and therefore the sensitivity, is bounded by T$_2^*$. The many-body gap protection demonstrated here extends this limit by nearly two orders of magnitude without refocusing pulses, offering a passive route to enhanced sensitivity in a solid-state device. With a superconducting cavity ($\kappa/2\pi\sim10$~kHz), the residual superradiant decay will be further suppressed, potentially extending T$_2^*$ toward the intrinsic T$_2>150$~ms\cite{Tiranov2025}.

Finally, the gap-protected Ramsey coherence may also benefit hybrid quantum interfaces. The optical transitions of $^{171}$Yb$^{3+}$ enable microwave--optical connectivity, and the extended T$_2^*$ will increase the storage time of spin-based quantum memories. More broadly, by demonstrating that cavity-mediated collective interactions can overcome the intrinsic disorder of a solid-state spin ensemble, these results establish rare-earth-doped crystals as a platform where many-body physics is an engineered resource rather than a limitation.

\begin{acknowledgments}
We acknowledge feedback on the manuscript from James Thompson and Ana Maria Rey. We acknowledge helpful discussions with Haoqing Zhang, Anjun Chu, and Emanuel Green. \textbf{Funding:} This work was supported by Institute of Quantum Information and Matter, an NSF Physics Frontiers Center (PHY-2317110) with support from the Moore foundation; U.S. Department of Energy, Office of Science, National Quantum Information Science Research Centers, Co-design Center for Quantum Advantage (C2QA) under contract number DE-SC0012704; Office of Naval Research grant no. N000142512278; and Gordon and Betty Moore Foundation, grant DOI 10.37807/GBMF12249. R.F. acknowledges the support from the Eddleman Graduate Fellowship. C.L. acknowledges support from the AWS Quantum Post-doctoral Fellowship. A.T. and P.G acknowledge support from Audace program CEA and the France 2030 program (QMemo, ANR-22-PETQ-0010). A.T. was supported by ANR under grant agreement no. ANR-22-CPJ2-0060-01. \textbf{Author Contributions:} R.F., C.L., and A.F. conceived the idea and experiment. R.F. and C.L. built the experimental setup, performed the measurements and simulations, and analyzed the data. A.T., K.W., and P.G. grew the sample. R.F., C.L., and A.F. wrote the manuscript with inputs from all authors. \textbf{Competing interests:} The authors declare no competing interests. \textbf{Data and materials availability:} All data are available in the manuscript or the supplementary information.
\end{acknowledgments}

\bibliography{references}
\clearpage
\onecolumngrid
\begin{center}
    \textbf{\Large{Supplementary Information}}
\end{center}

\renewcommand{\theequation}{S\arabic{equation}}
\setcounter{equation}{0}
\renewcommand{\thesection}{S\arabic{section}}
\setcounter{section}{0}
\renewcommand{\thetable}{S\arabic{table}}
\setcounter{table}{0}
\renewcommand{\thefigure}{S\arabic{figure}}
\setcounter{figure}{0}

\section{Properties of $\mathbf{{}^{171}\mathrm{Yb}^{3+}}$ in $\mathbf{\mathrm{CaWO}_4}$}

\begin{figure}[htbp]
\centering\includegraphics[width=0.6\textwidth]{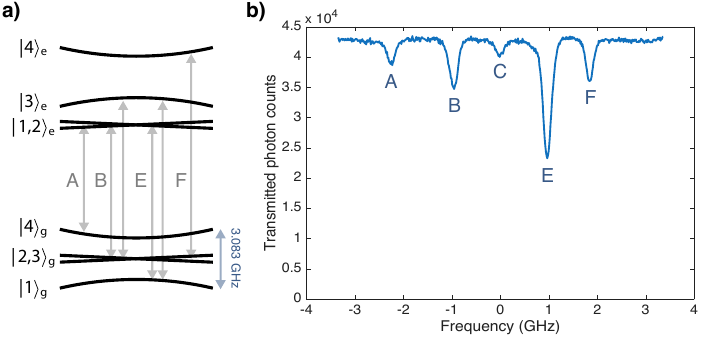}
\caption{\textbf{${}^{171}\mathrm{Yb}^{3+}:\mathrm{CaWO}_4$ energy levels.} \textbf{(a)} Energy levels, with labeled optical transitions accessible by light polarized along the crystal $a$-axis. \textbf{(b)} Optical absorption spectrum with the crystal fully thermalized, with an extracted spin temperature of 80~mK. The C line arises from even-mass Yb isotopes with no hyperfine structure. Both the B and E lines each comprise two transitions,  the splitting between $\lvert{1,2}\rangle_e$ and $\lvert{3}\rangle_e$ is smaller than the optical inhomogeneous linewidth and is therefore unresolved at zero magnetic field.}
\label{figS:levels}
\end{figure}

$\mathrm{CaWO}_4$ is a scheelite-structure crystal in which $\mathrm{Yb}^{3+}$ substitutes for $\mathrm{Ca}^{2+}$ at a site with $S_4$ point symmetry. The crystal used in this work is isotopically enriched with $^{171}\mathrm{Yb}$ to a total $^{171}\mathrm{Yb}^{3+}$ concentration of 4.96~ppm at the relevant tetragonal $S_4$ site\cite{Tiranov2025}. The remaining even-mass Yb isotopes ($^{170,172,174,176}\mathrm{Yb}$, $I=0$) have no hyperfine structure; their unsplit transition gives rise to the C transition visible in the optical spectrum (Fig.~\ref{figS:levels}b).

The optical and spin properties of ${}^{171}\mathrm{Yb}^{3+}:\mathrm{CaWO}_4$ have been characterized in detail in Tiranov et al.~\cite{Tiranov2025}. $^{171}\mathrm{Yb}^{3+}$ ($I=1/2$) has a Kramers doublet ground state ${}^2\mathrm{F}_{7/2}(0)$ whose degeneracy is lifted by the hyperfine interaction into four levels. At zero magnetic field, the electron spin ($S=1/2$) and nuclear spin ($I=1/2$) are strongly hybridized, yielding the eigenstates:
\begin{align}
    \lvert{1,4}\rangle_g &= \left( \lvert{\uparrow_s \downarrow_n}\rangle_g \pm \lvert{\downarrow_s \uparrow_n}\rangle_g \right)/\sqrt{2}\\
    \lvert{2,3}\rangle_g &= \lvert{\uparrow_s \uparrow_n}\rangle_g, \lvert{\downarrow_s \downarrow_n}\rangle_g
\end{align}
where $\lvert{\uparrow_s,\downarrow_s}\rangle = \lvert{S=1/2, m_S=\pm 1/2}\rangle$ and $\lvert{\uparrow_n,\downarrow_n}\rangle = \lvert{I=1/2, m_I=\pm 1/2}\rangle$. The spin qubit transition $\lvert{1}\rangle_g \leftrightarrow \lvert{4}\rangle_g$ at $\omega_s/2\pi = 3.08385$~GHz is a zero first-order Zeeman (ZEFOZ) transition: because both states are equal superpositions of $m_S = \pm 1/2$, the expectation value of the magnetic moment vanishes, making the transition frequency insensitive to magnetic field fluctuations to first order. This gives rise to the narrow inhomogeneous linewidth $\gamma_\mathrm{inh}/2\pi < 5$~kHz and long spin-echo coherence time T$_2 > 150$~ms\cite{Tiranov2025}. Despite the first-order magnetic insensitivity, the transition retains a substantial magnetic dipole matrix element ($g_\parallel = 1.08$) because it connects states with different symmetry under spin exchange, enabling strong coupling to the resonator magnetic field.

Similarly, the optical excited state ${}^2\mathrm{F}_{5/2}(0)$ splits into:
\begin{align}
    \lvert{1,2}\rangle_e &= \lvert{\uparrow_s \uparrow_n}\rangle_e, \lvert{\downarrow_s \downarrow_n}\rangle_e\\
    \lvert{3,4}\rangle_e &= \left( \lvert{\uparrow_s \downarrow_n}\rangle_e \pm \lvert{\downarrow_s \uparrow_n}\rangle_e \right)/\sqrt{2}
\end{align}

The optical transitions connecting the ground and excited manifolds are labeled A--F, where we show only the transitions accessible via light polarized perpendicular to the crystal $c$-axis ($E \perp c$) (Fig.~\ref{figS:levels}a). In our experiment, the A transition is used for optical pumping out of $\lvert{4}\rangle_g$ ($\equiv\lvert{\downarrow}\rangle$), while an electro-optic modulator sideband addresses the F transition to clear residual population in $\lvert{2,3}\rangle_g$. Population readout is performed by sweeping a probe laser across the entire absorption spectrum and measuring the absorption depths on the A and E transitions (and ignoring C), which are proportional to the $\lvert{\downarrow}\rangle$ and $\lvert{\uparrow}\rangle$ populations, respectively.

\section{Experimental setup}
The CaWO$_4$ crystal is a $4.4$~mm $\times$ $4.6$~mm $\times$ $5$~mm cuboid ($a \times a \times c$) with both $c$-cut faces polished for optical transmission. The $^{171}$Yb$^{3+}$ concentration at the relevant tetragonal S4 site is 4.96~ppm\cite{Tiranov2025}. The crystal is glued and thermally anchored to the center of the loop-gap resonator with low-temperature GE varnish. The resonator is mounted between two Thorlabs FiberPort collimators for optical pumping and transmission measurements. The optical coupling efficiency is approximately $30$--$50$~\%, varying between cooldowns. The setup is placed on the mixing chamber plate of a Bluefors dilution refrigerator (LD400) with a base temperature below 30~mK. Two separate loop-gap resonators were used: an unpolished, overcoupled resonator ($\kappa/2\pi \approx 4$~MHz) for superradiance measurements (Fig.~2), and a polished, undercoupled resonator ($\kappa/2\pi = 660$~kHz) for dispersive measurements (Figs.~1, 3, 4). The broader linewidth of the former facilitates tuning onto the spin resonance at cryogenic temperatures, while the narrower linewidth of the latter maximizes the ratio of coherent interaction to dissipation. The same crystal was used in both resonators.

A $973$~nm external cavity diode laser (Toptica DL Pro) is used for optical pumping. The center frequency is locked to a stable reference cavity (Stable Laser Systems) at the A transition for pumping population out of $\lvert{4}\rangle_g$. A fiber-coupled electro-optic phase modulator (Exail MPZ-LN-10) is driven at $3.97$~GHz to create a sideband on the F transition for clearing population out of $\lvert{2}\rangle_g$ and $\lvert{3}\rangle_g$. A fiber-coupled acousto-optical modulator (Aerodiode 940AOM-2) gates the optical pumping pulse. The laser power for optical pumping is typically $\sim5$~mW. A fiber-coupled optical switch (Agiltron FFSM-12LC01133) alternates between the pump laser and a second $973$~nm laser (Toptica CTL) for absorption spectroscopy. The CTL laser is swept $\pm3$~GHz around the C transition to cover all optical transitions for population readout. Transmitted photons are detected by a superconducting nanowire single-photon detector (Photonspot) at 100~mK, with photon counts time-tagged (Swabian Time Tagger 20).

A microwave frequency source (Rohde \& Schwarz, SGS100A) drives the spin transition. For agile phase control, the source output (at $\omega_s+300$~MHz) is sent to the LO port of an IQ mixer (Marki Microwave, IQ-1545LMP) and mixed with two RF inputs (at 300~MHz) from an arbitrary waveform generator (Zurich Instruments, HDAWG). The relative phase, amplitude and offset are tuned to produce a single signal sideband at the spin frequency. This signal is gated by two cascaded RF switches (ZASWA-2-50DR+) and amplified (ZHL-16W-43-S+) before entering the resonator. In the superradiance measurements, one resonator port is used for excitation and the other for collecting the emitted microwave field. The output near the spin frequency is mixed down to $20$~MHz and digitized with an analog-to-digital converter (AlazarTech ATS9364).

\section{Loop-gap resonator field simulation}
The resonator's resonance frequency and field distribution were simulated using COMSOL's emw package. The magnetic field along the crystal $c$-axis is approximately uniform across the crystal volume: 90\% of sites are within 5\% of the field maximum, with a standard deviation of only 2\%. This applies to both resonators used in this work as the geometry is identical. The single-spin coupling is
\begin{equation}
    g=\frac{g_{\parallel}\mu_B}{2\hbar}\sqrt{\frac{\mu_0\hbar\omega_s}{2V_m}},
\end{equation}
where $g_\parallel=1.08$, $\mu_B$ is the Bohr magneton, $\mu_0$ is the vacuum permeability, and $V_m=275$~mm$^3$ is the simulated mode volume, giving $g/2\pi\approx15$~mHz.

\begin{figure}[htbp]
\centering\includegraphics[width=0.8\textwidth]{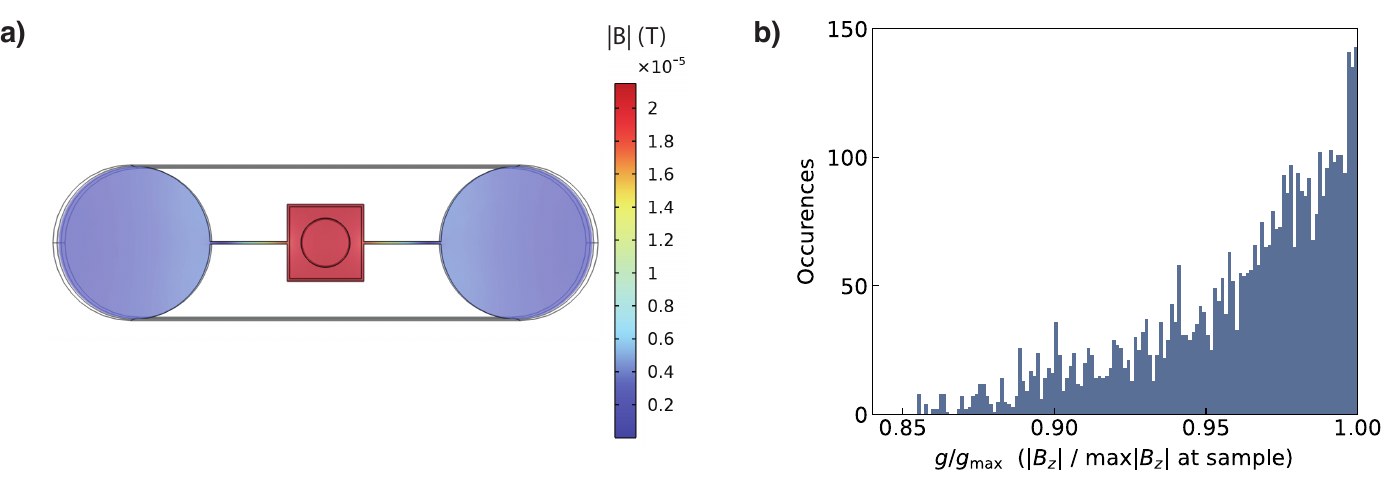}
\caption{\textbf{Resonator simulations.} \textbf{(a)} COMSOL simulation of the loop-gap resonator's magnetic field along the $z$-axis (into the page). The uniformly colored central square houses the crystal.
    \textbf{(b)} Histogram of the normalized magnetic field sampled across the crystal volume. This is equivalent to the normalized ion--cavity coupling $g/g_\text{max}$ of individual ions.}
\label{figS:resonator}
\end{figure}

\section{Spin diffusion during state initialization}

In our experiment, the initial states of the ions are polarized through optical pumping with a beam covering $\sim$1\% of the crystal cross section. Nevertheless, the resonator $S_{21}$ shows that the collective coupling continues to increase with longer initialization time, well beyond the $\sim$1~s timescale for polarizing the optically addressed ions. This reveals a spin-diffusion mechanism\cite{Welinski2020} that gradually redistributes polarization to ions outside the optical beam.

The collective coupling is extracted by fitting the measured $S_{21}$ to the transfer function:
\begin{equation}
    S_{21}\left(\omega\right) = \frac{i \kappa_{\text{out}}}{\omega - \omega_c + i\kappa/2 - \frac{(g\sqrt{N_0})^2}{\omega-\omega_{s} + i \gamma_{\text{inh}}/2}},
    \label{eqn:S21}
\end{equation}
where $\kappa_{\text{out}}$ is the external coupling rate. The measured $S_{21}$ at each optical pumping time is fit to Eq.~(\ref{eqn:S21}) to extract $g\sqrt{N_0}$ as shown in Fig.~\ref{fig:fig1}g.

\section{Theoretical framework and numerical simulations}

\subsection{Effective Hamiltonian}

We start with the Tavis-Cummings model describing the coupling between the spin ensemble and resonator in the rotating frame of the spin frequency $\omega_s$:
\begin{equation}
    \hat{H}=\Delta \hat{a}^{\dagger}\hat{a} + \frac{1}{2}\sum_{j=1}^{N}\delta_j\hat{\sigma}_{z,j}+g\sum_{j=1}^N(\hat{a}^{\dagger}\hat{\sigma}_{-,j}+\hat{\sigma}_{+,j}\hat{a}),
\end{equation}
where $\hat{a}^\dagger (\hat{a})$ is the creation (annihilation) operator of the cavity mode, $\hat{\sigma}_{z,j},\hat{\sigma}_{\pm,j}$ are the Pauli operators for the $j^{th}$ spin, $\Delta=\omega_c-\omega_s$ is the resonator--spin detuning, $\delta_j$ is the frequency shift of the $j^{th}$ spin from the mean spin transition, distributed with zero mean and full-width at half maximum $\gamma_{\mathrm{inh}}$, and $g$ is the single spin--resonator coupling strength, which we approximate to be uniform across the whole ensemble. The photon loss from the resonator with power decay rate $\kappa$ is $\sqrt{\kappa} a$.

For the experiments in the dispersive regime, we are in the limit $\Delta \gg \sqrt{N} g$ and $\Delta \gg \kappa$ ( $\Delta/g\sqrt{N_0} \gtrsim 60$ and $\Delta/\kappa \approx 33$), and the cavity field can be adiabatically eliminated, leading to a steady-state cavity response determined by the spin dynamics
\begin{equation}
    \hat{a}(t) = \frac{i g}{i \Delta + \kappa/2} \sum_{j=1}^N \hat{\sigma}_{-,j},
\end{equation}
which gives rise to the effective spin-spin interaction Hamiltonian\cite{Norcia2018} $\hat{H}_{eff} = \sum_{i,j} \chi \hat{\sigma}_{+,i} \hat{\sigma}_{-,j} + \sum_j \delta_j \hat{\sigma}_{z,j} $ with the collective jump operator $\hat{L}_{cav} = \sqrt{\Gamma_\text{SR}/2} \sum_j \hat{\sigma}_{-,j}$. Here, $\chi = 4g^2 \Delta / \left( 4 \Delta^2 + \kappa^2 \right) $ and $\Gamma_\text{SR} = 4 g^2 \kappa / \left( 4 \Delta^2 + \kappa^2 \right) $ are the characteristic rates of the cavity-induced interaction and dissipation for single particles.

Together with single-particle dephasing rate $\gamma_2 = 2\pi/\mathrm{T}_2$, the effective Hamiltonian and jump operator for the whole system are
\begin{align}
    \hat{H}_{eff} &= \chi \hat{J}_+ \hat{J}_- + \sum_j \delta_j \hat{\sigma}_{z,j} \\
    \hat{L} &= \sqrt{\Gamma_\text{SR}/2} \hat{J}_- + \sqrt{\gamma_2} \sum_j \hat{\sigma}_{z,j},
\end{align}
where $\hat{J}_{\pm,z} = \frac{1}{2}\sum_j \hat{\sigma}_{(\pm,z),j}$ are the collective spin operators. A complementary derivation using second-order perturbation theory can be found in Luo et al.~\cite{Thompson2024Science}.

The effective cavity-mediated spin-exchange interaction expands as $\chi \hat{J}_+ \hat{J}_- = \chi \left( \hat{J}^2 - \hat{J}_z^2 +\hat{J}_z\right)$. The $\chi \hat{J}^2$ term gives rise to a many-body energy gap between collective states with different symmetry. The $\chi \hat{J}_z^2$ term gives rise to the one-axis-twisting dynamics. In the large ion number limit, the single particle rotation term $\chi \hat{J}_z$ can be neglected.

\subsection{Mean-field equations of motion}

At finite ion-cavity detuning, with the effective Hamiltonian and jump operator, the equations of motion for the $j^{th}$ ion in the Heisenberg picture are:
\begin{align}
    \frac{d\hat{\sigma}_{+,j}}{dt}  & = -2 i \chi \hat{J}_+ \hat{\sigma}_{z,j} + i \delta_j \hat{\sigma}_{+,j}
    +\frac{\Gamma_\text{SR}}{2} \hat{J}_+ \hat{\sigma}_{z,j} - \gamma_2 \hat{\sigma}_{+,j} \\
    \frac{d\hat{\sigma}_{z,j}}{dt} & = i \chi \left[ \hat{J}_+ \hat{\sigma}_{-,j}
     - \hat{J}_- \hat{\sigma}_{+,j} - 2\hat{\sigma}_{z,j}\right]
     -\frac{\Gamma_\text{SR}}{4} \left[  \hat{J}_+ \hat{\sigma}_{-,j}  +\hat{J}_- \hat{\sigma}_{+,j} + \hat{\sigma}_{z,j}  \right]
\end{align}

When the cavity is on resonance with the spin ensemble, in the limit where cavity linewidth is larger than all other frequency scales, the spin dynamics is dominated by superradiant decay with the equations of motion of the collective spin:
\begin{align}
    \frac{d\hat{J}_+}{dt}  & = +\frac{\Gamma_\text{SR}}{2} \hat{J}_+ \hat{J}_z \\
    \frac{d\hat{J}_z}{dt} & = 
     -\frac{\Gamma_\text{SR}}{4} \left[\hat{J}_+ \hat{J}_-  +\hat{J}_- \hat{J}_+ + \hat{J}_z  \right]
\end{align}
The initial state is a coherent spin state at polar angle $\theta$ from the south pole: $\hat{J}_-(0) = -\sin\theta/2$, $\hat{J}_z(0) = -\cos\theta/2$. The emitted field intensity is proportional to $\langle J_+ J_- \rangle$.

The superradiant emission traces in Figs.~2c and 2d are modeled by numerically integrating these equations using an implicit Runge--Kutta method. For each optical pumping duration (equivalently varying $g\sqrt{N_0}$), the superradiant emission traces are simulated using independently calibrated experimental parameters.

\subsection{Ramsey coherence simulations}
We simulate the Ramsey coherence decay by numerically integrating the mean-field equations of motion (Eqs.~S11--S12) for $N_{\mathrm{sim}}=10{,}000$ ion groups. Each group is assigned a frequency offset $\delta_j$ drawn from a distribution with FWHM $\gamma_{\mathrm{inh}}/2\pi = 5$~kHz, and the initial state is on the equator along $\hat{x}$. The residual coherence after delay $\tau$ is given by the normalized transverse magnetization $2\lvert\langle\hat{J}_\perp\rangle\rvert/N_0$.

The spectroscopic measurement constrains the inhomogeneous linewidth to  a FWHM of order $\gamma_{\mathrm{inh}}/2\pi \approx 4$--$6$~kHz, but  does not uniquely determine the lineshape. Because the Ramsey envelope  depends on the underlying distribution, there is no unique conversion between the measured spectral FWHM and the low-interaction Ramsey time $\mathrm{T}_2^*$. For the same FWHM, Gaussian, Lorentzian, and intermediate Voigt profiles produce different decay envelopes and differ  in the corresponding $1/e$ coherence time by order-unity factors. We  therefore use $\gamma_{\mathrm{inh}}$ as the scale of the static disorder  and compare the data to simulations spanning Gaussian, Lorentzian, and  intermediate Voigt distributions.

\begin{figure}
\centering\includegraphics[width=\textwidth]{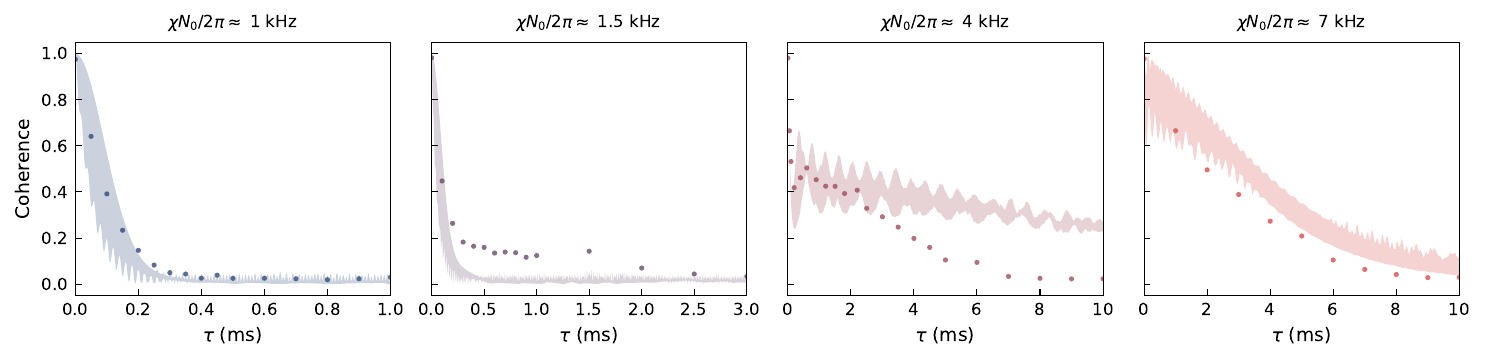}
\caption{Ramsey coherence decay as a function of delay time $\tau$ for four values of $\chi N_0/2\pi$. Data are shown as colored markers; mean-field simulations are shown as shaded bands reflecting uncertainties in the inhomogeneous lineshape (Gaussian to Lorentzian), linewidth ($\gamma_{\mathrm{inh}}/2\pi = 4$--$6$~kHz), and interaction strength ($\pm 20$--$30\%$ in $\chi N_0$).}
\label{figS:ramsey}
\end{figure}

\begin{figure}
\centering\includegraphics[width=0.4\textwidth]{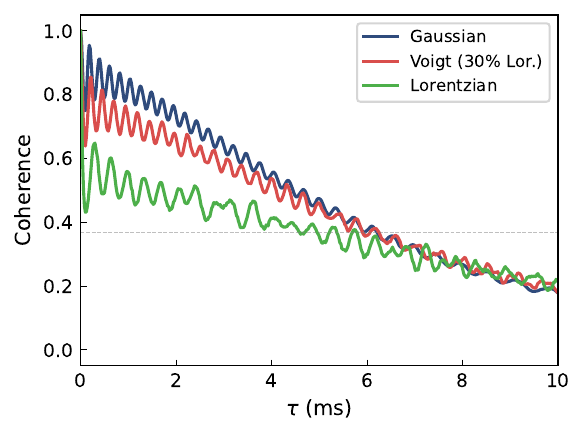}
\caption{Sensitivity of the simulated coherence decay to the inhomogeneous lineshape at $g\sqrt{N_0}/2\pi = 350$~kHz. Gaussian (blue), Voigt with 30\% Lorentzian fraction (orange), and pure Lorentzian (green) distributions are compared, all with the same FWHM of 5~kHz.}
\label{figS:lineshape}
\end{figure}

The simulations reproduce the key experimental phenomenology  (Fig.~\ref{figS:ramsey}): rapid inhomogeneous dephasing at low $\chi N_0$, emergence of a two-component (double-exponential) decay at intermediate  $\chi N_0 \sim \gamma_{\mathrm{inh}}$, and sustained coherence at high $\chi N_0$. 
%At intermediate interaction strengths, spins with frequency offsets $|\delta_j| \lesssim \chi N_0$ are energetically protected by the many-body gap, while those with $|\delta_j| \gtrsim \chi N_0$ dephase rapidly, producing the observed fast-then-slow decay dynamics.
The quantitative agreement between simulation and experiment depends on the assumed shape of the inhomogeneous distribution. A Gaussian distribution overestimates the coherence at long times, while a Lorentzian distribution underestimates it. The experimental data largely fall between these bounds, consistent with a Voigt profile reflecting contributions from both intrinsic crystal field disorder and Lorentzian broadening from superhyperfine coupling to neighboring nuclear spins or residual magnetic fields (Fig.~\ref{figS:lineshape}).

\end{document}